\documentclass[usenatbib]{mnras}

\usepackage{graphicx}
\usepackage{bm}
\usepackage{amssymb,amsmath}
\usepackage{mathrsfs} \usepackage{latexsym}
\usepackage{mathtools} \usepackage{color}
\usepackage[normalem]{ulem} 
\usepackage{dcolumn}
\hypersetup{dvips, colorlinks=true, linkcolor=black, citecolor=black, filecolor=blue, urlcolor=black}
\usepackage[usenames,dvipsnames]{xcolor}

\allowdisplaybreaks

\usepackage{soul} \usepackage[dvipsnames]{xcolor}
\definecolor{aquamarine}{rgb}{0.5, 1.0, 0.83}

\newcommand{\p}{\partial}
\newcommand{\rd}{\mathrm{d}}
\newcommand{\re}{\mathrm{e}}
\newcommand{\ri}{\mathrm{i}}
\newcommand{\MBH}{M_{\bullet}}
\newcommand{\bK}{\mathbf{K}}
\newcommand{\bKp}{\mathbf{K}^{\prime}}
\newcommand{\hbL}{\widehat{\mathbf{L}}}
\newcommand{\rh}{\mathrm{h}}
\newcommand{\yr}{\mathrm{yr}}
\newcommand{\kyr}{\mathrm{kyr}}
\newcommand{\Myr}{\mathrm{Myr}}
\newcommand{\kpc}{\mathrm{kpc}}
\newcommand{\mpc}{\mathrm{mpc}}
\newcommand{\Tc}{T_{\mathrm{c}}}
\newcommand{\mJ}{\mathcal{J}}
\newcommand{\bO}{\mathbf{\Omega}}
\newcommand{\bD}{\mathbf{D}}
\newcommand{\half}{\tfrac{1}{2}}
\newcommand{\erf}{\mathrm{erf}}
\newcommand{\Msun}{M_{\odot}}
\newcommand{\deltaD}{\delta_{\mathrm{D}}}
\newcommand{\bP}{\mathbf{P}}
\newcommand{\bA}{\mathbf{A}}
\newcommand{\bM}{\mathbf{M}}
\newcommand{\bR}{\mathbf{R}}
\newcommand{\bI}{\mathbf{I}}
\newcommand{\bU}{\mathbf{U}}

\newcommand{\Diag}{\mathrm{Diag}}
\newcommand{\phicut}{\phi_{\mathrm{cut}}}
\newcommand{\ellmax}{\ell_{\mathrm{max}}}

\newcommand{\tmax}{t_{\mathrm{max}}}
\newcommand{\mO}{\mathcal{O}}
\newcommand{\Max}{\mathrm{Max}}
\newcommand{\Min}{\mathrm{Min}}
\newcommand{\bT}{\boldsymbol{\theta}}
\newcommand{\tmu}{\widetilde{\mu}}
\newcommand{\tsigma}{\widetilde{\sigma}}
\newcommand{\bone}{\mathbf{1}}
\newcommand{\RePart}{\mathrm{Re}}

\usepackage{acronym}
\newacro{BH}{black hole}
\newacroplural{BH}[BHs]{black holes}
\newcommand{\BH}{\ac{BH}}

\newacro{DF}{distribution function}
\newacroplural{DF}[DFs]{distribution functions}
\newcommand{\DF}{\ac{DF}}

\newacro{PDF}{probability distribution function}
\newacroplural{PDF}[PDFs]{probability distribution functions}
\newcommand{\PDF}{\ac{PDF}}

\newacro{IMBH}{intermediate mass black hole}
\newacroplural{IMBH}[IMBHs]{intermediate mass black holes}
\newcommand{\IMBH}{\ac{IMBH}}
\newcommand{\IMBHs}{\acp{IMBH}}
\newacro{VRR}{vector resonant relaxation}
\newcommand{\VRR}{\ac{VRR}}

\begin{document}

\title[Constraining IMBHs from SgrA*'s disc]{Constraining intermediate-mass black holes from the stellar disc of SgrA*}

\author[J.-B. Fouvry, M.~J. Bustamante-Rosell, A. Zimmerman]{Jean-Baptiste Fouvry$^{1}$, Mar\'{i}a Jos\'{e} Bustamante-Rosell$^{2}$, Aaron Zimmerman$^{3}$\\
\noindent
$^{1}$ CNRS and Sorbonne Universit\'e, UMR 7095, Institut d'Astrophysique de Paris, 98 bis Boulevard Arago, F-75014 Paris, France\\
$^{2}$ Department of Astronomy and Astrophysics, University of California, Santa Cruz, CA 95064, USA\\
$^{3}$ Weinberg Institute, University of Texas at Austin, Austin, TX 78712, USA
}

\maketitle

\begin{abstract}
Stars evolving around a supermassive black hole
see their orbital orientations diffuse
efficiently, a process called ``vector resonant relaxation''.
In particular, stars within
the same disc, i.e.\ neighbors in orientations,
will slowly diffuse away from one another
through this stochastic process.
We use jointly (i) detailed kinetic predictions for the
efficiency of this dilution
and (ii) the recent observation of a stellar disc around SgrA*,
the supermassive black hole at the centre of the Milky-Way,
to constrain SgrA*'s unobserved stellar cluster.
Notably, we investigate quantitatively the impact of a population
of intermediate mass black holes on the survivability of the stellar disc.
\end{abstract}


\section{Introduction}
\label{sec:Introduction}

Recent outstanding observations keep providing us with
new information on the dense cluster orbiting around SgrA*,
the supermassive \BH\ at the centre of the Milky-Way.
These include a thorough census of stellar populations~\citep{Ghez+2008,Gillessen+2017}
in particular highlighting the presence of a clockwise stellar disc~\citep{Levin+2003,Paumard+2006,Bartko+2009,Lu+2009,Yelda+2014,vonFellenberg+2022},
along with the relativistic precession of the star S2~\citep{Gravity+2020}.
This wealth of information regarding the dynamical status of SgrA*'s
nuclear stellar cluster~\citep{Neumayer+2020}
is expected to offer new insights on the possible presence
of \IMBHs\ in these dense regions~\citep[see, e.g.\@,][]{PortegiesZwart+2002}.

The dynamics of stars orbiting a supermassive \BH\ involves
an intricate hierarchy of dynamical processes~\citep{Rauch+1996,Alexander2017}.
In the present work, we focus on the process of \VRR\@~\citep{Kocsis+2015},
i.e.\ the mechanism through which coherent torques between orbits
lead to an efficient diffusion of the stellar orbital orientations.
The analytical and numerical studies of \VRR\ have recently seen a surge
of new developments
to understand in particular
(i) the warping of stellar discs~\citep{Kocsis+2011};
(ii) the enhancement of binary merger rates embedded
in galactic nuclei~\citep{Hamers+2018};
(iii) the non-trivial equilibrium distribution of orientations~\citep{Roupas+2017,Takacs+2018,Touma+2019,Gruzinov+2020,Magnan+2022,Mathe+2023} 
possibly leading to formation of discs of \IMBHs\@~\citep{Szolgyen+2018};
(iv) the rapid (resonant) dynamical friction imposed by a stellar disc
onto an \IMBH\@~\citep{Ginat+2022,Levin2022}.

Similarly, the potential fluctuations
generated by the old background stellar cluster and its putative \IMBHs\
can drive the spontaneous dilution of the presently observed
discs of young S-stars. As such, we argue that the present survival
of the stellar disc 
requires that \VRR\ fluctuations are not driven
too strongly throughout the disc's life.
In the present work, we build upon~\cite{Giral+2020},
hereafter~{G20},
and use quantitatively the process of ``neighbor separation'' by \VRR\@
to compare \VRR\ predictions to the observed
S-stars.
Our aim is to leverage \VRR\
to constrain the stellar content of SgrA*'s cluster,
in particular its possible \IMBHs\@.

The paper is organised as follows.
In \S\ref{sec:VRR}, we recall the key elements
of \VRR\@. In \S\ref{sec:Inference}, we detail
our inference method. In \S\ref{sec:Results},
we apply this method to SgrA* to (loosely) constrain
the properties of the likely population of \IMBHs\
orbiting therein. Finally,
we conclude in \S\ref{sec:Conclusions}.
The calculations and numerical details
presented in the main text are kept to a minimum
and are spread through Appendices.

\section{Vector resonant relaxation}
\label{sec:VRR}

We consider a set of ${ N \!\gg\! 1 }$ stars orbiting 
around a supermassive \BH\ of mass ${ \MBH }$.
We call this background of stars the ``bath''. They represent
the unobserved populations (old low mass stars, \IMBHs\@, etc.).
They are responsible for fluctuations in the gravitational potential,
and hence drive \VRR\@. 
Provided that one considers dynamics on timescales longer
than the Keplerian motion induced by the central \BH\
(e.g.\@, ${ \sim 16 \, \yr }$ for S2) and the in-plane precession
driven by relativistic corrections (e.g.\@, ${ \sim 3 \!\times\! 10^{4} \, \yr}$ for S2),
one can orbit-average the dynamics over each star's mean anomaly
and pericentre phase. Stars are formally replaced by massive annuli,
see, e.g.\@, fig.~{1} in~{G20}. These are characterised by
\begin{equation}
\bK = (m , a, e) ,
\label{def_bK}
\end{equation}
with $m$ the individual mass,
$a$ the semi-major axis,
and $e$ the eccentricity. The norm of the angular momentum of a given annulus
is ${ L (\bK) \!=\! m \sqrt{G \MBH a (1 \!-\! e^{2})} }$
and is conserved during the \VRR\ dynamics.
The remaining dynamical quantity
is the instantaneous orbital orientation,
denoted with the unit vector ${ \hbL }$.
We assume that the bath is on average isotropic.
Its distribution of orbital parameters is characterised
by a \DF\@, ${ n (\bK) }$, normalised to
${ \!\int\! \rd \bK \, \rd \hbL \, n (\bK) \!=\! N }$.
Our conventions are spelled out in~\S\ref{sec:StellarCusps}.

\begin{figure}
\centering
\includegraphics[width=0.35 \textwidth]{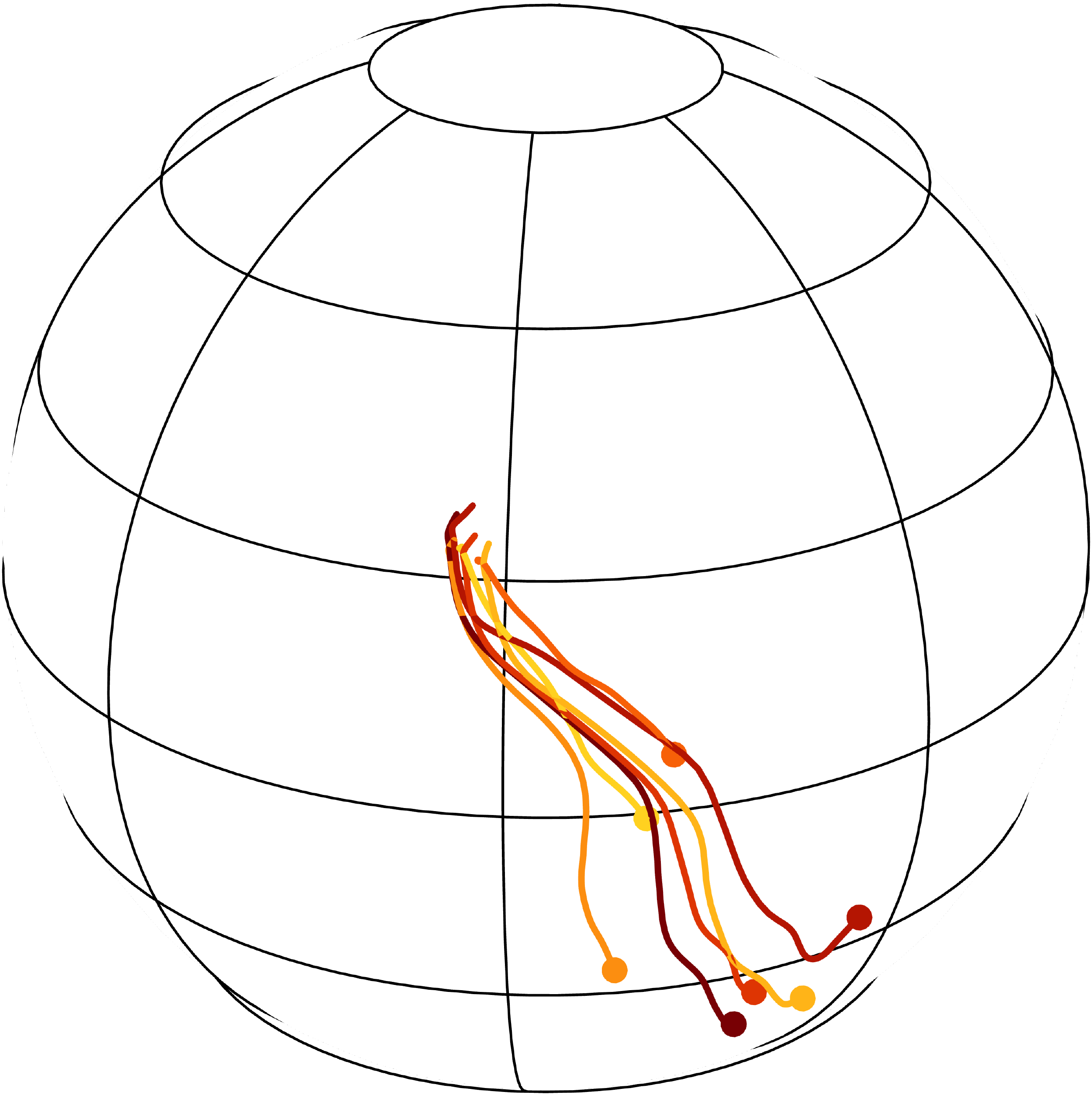}
\caption{Illustration of the \VRR\ dilution of a disc of 7 stars
over ${3 \, \Myr}$ when embedded within the Top-Heavy model
from~\S\ref{sec:LessSimpleFiducial}.
Here, each point represents the instantaneous orbital orientation, $\hbL$,
of a star on the unit sphere.
}
\label{fig:Dilution}
\end{figure}

In addition to this bath, we consider a population
of test stars.
These correspond to the observed S-stars forming the clockwise stellar disc.
We neglect the mass of these test stars: they only probe
the potential fluctuations in the system but do not contribute to it.
Ultimately, these stars' orientations are driven away from one another via \VRR\@,
i.e.\ the disc dissolves as illustrated in Fig.~\ref{fig:Dilution}.
This is the observational signature
that we leverage in order to constrain
the stellar content of the background bath, i.e.\ ${ n (\bK) }$.

Given that the disc stars are massless,
investigating the dilution of the disc is formally equivalent
to investigating the dilution of pairs of test stars~(see~\S{E} in~{G20}).
Let us consider two disc stars
of orbital parameters $\bK_{1}$ and $\bK_{2}$. We assume that at the time
of their birth, these stars' respective angular momentum vectors
were separated by some small angle $\phi_{0}$. As a result of \VRR\@,
these two stars are slowly stirred away from one another.
This is the process of ``neighbor separation''
driving the stochastic evolution of ${ \phi (t) }$.
This was the process investigated in detail in~{G20}.

In~\S\ref{sec:NeighSep}, we reproduce and improve upon
the main results of~{G20}. Ultimately, \S\ref{sec:NeighSep}
provides us with a prediction for the time evolution of the \PDF\
of the angular separation
\begin{equation}
t \mapsto P (\phi \,|\, t) ,
\label{pred_Pphi}
\end{equation}
as a function of time.
We emphasise that the statistics
of ${ P (\phi \,|\, t) }$ depends
on
(i) the orbital distribution of the background bath, ${ n (\bK) }$;
(ii) the orbital parameters of the two test stars, ${ (\bK_{1} , \bK_{2}) }$;
(iii) their initial angular separation, $\phi_{0}$.

\section{Inference methods}
\label{sec:Inference}

Since \VRR\ drives a stochastic dynamics,
constraints on the properties of SgrA*'s stellar background,
i.e.\ ${ n (\bK) }$, can only be obtained in a statistical sense.
To do so, we use maximum likelihood estimations.
We detail our approach in~\S\ref{sec:Likelihood}
and reproduce here only the key assumptions.

Our goal is to constrain the parameters, $\bT$,
of SgrA*'s background bath, given some observed
angular separations between disc stars.
For example, parameters
can be the collection ${ \bT \!=\! \{ n (\bK) , \phi_{0} , T_{\star} \} }$,
with $T_{\star}$ the ages of the observed stars.
For a given pair of test stars ${ (i,j) }$,
we have at our disposal their (conserved) orbital parameters, ${ \{ \bK_{i} , \bK_{j} \} }$,
along with their current angular separation, ${ \phi_{ij} }$.
For each $\bT$, we compute the associated predicted \PDF\@,
${ \phi_{ij} \!\mapsto\! P (\phi_{ij} \,|\, \bT, \bK_{i} , \bK_{j}) }$,
according to which $\phi_{ij}$
could have been drawn.
The likelihood of the parameters $\bT$ is then proportional to
${P (\phi_{ij} \,|\, \bT, \bK_{i} , \bK_{j}) }$.
This is the gist of our method.

To improve upon this approach,
we include a few additional effects by
(i) marginalising over the dispersion of the predicted
log-normal distributions 
which we use to model the likelihood
(\S\ref{sec:MarginalSigma});
(ii) accounting jointly for all the distinct pairs
of test stars of a given disc (\S\ref{sec:DiscDilution});
(iii) incorporating measurement errors
in the observed S-stars (\S\ref{sec:MeasurementErrors}).
In practice, all these effects are included
by bootstrapping over independent samples.
It is essential to assess the performance of this approach
by applying it to tailored numerical simulations
of increasing complexity, see~\S\ref{sec:NumericalSimulations}
for details. This is what we now detail.

\begin{figure}
\centering
\includegraphics[width=0.45 \textwidth]{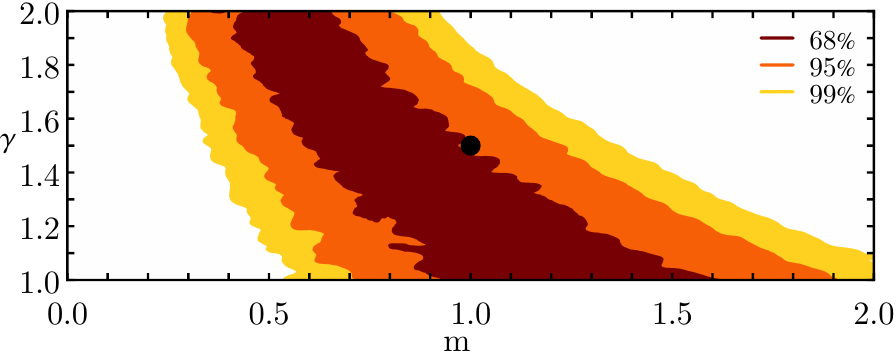}\\
\includegraphics[width=0.45 \textwidth]{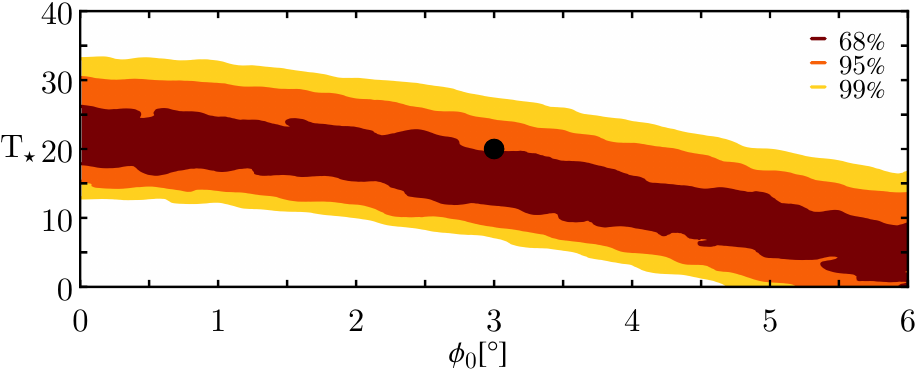}
\caption{Likelihood for the fiducial one-population simulations
(\S\ref{sec:SimpleFiducial})
as one varies ${ (m , \gamma) }$ (top) or ${ (\phi_{0} , T_{\star}) }$ (bottom).
The black dot corresponds to the bath effectively simulated
and the colored lines to the confidence levels.
See~\S\ref{sec:MarginalSigma} for details.
}
\label{fig:CornerFiducial}
\end{figure}

First, we consider a simple one-population background bath
(\S\ref{sec:SimpleFiducial}).
In Fig.~\ref{fig:CornerFiducial}, we report our 
inferred likelihoods
from these fiducial simulations,
as one varies respectively ${ (m,\gamma) }$,
the individual mass and power-law index
of the bath, or ${ (\phi_{0} , T_{\star}) }$,
the initial angular separation and age of the test stars.
The likelihood here corresponds to synthetic observations
mimicking our actual analysis of the S-stars around SgrA*.
In fact, we display the expectation value of the likelihood over multiple 
realisations of the observations, in order to provide a stringent test of the model,
as detailed in~\S\ref{sec:SimpleFiducial}.
Although Fig.~\ref{fig:CornerFiducial} suffers from a slight bias,
our method is able to recover the parameters
of the simulated bath. This is reassuring.

\begin{figure}
\centering
\includegraphics[width=0.45 \textwidth]{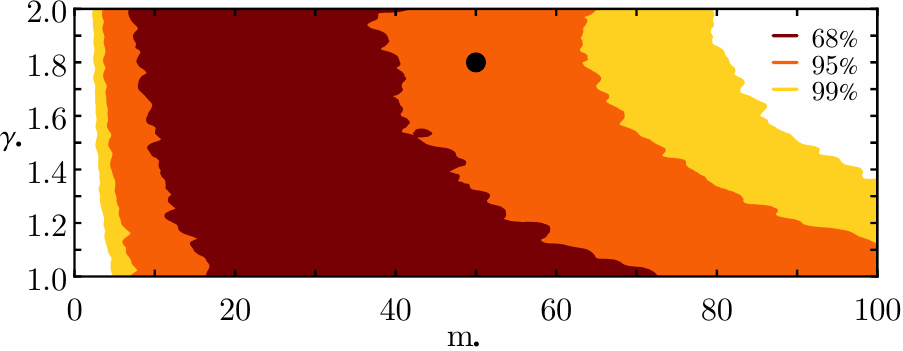}\\
\includegraphics[width=0.45 \textwidth]{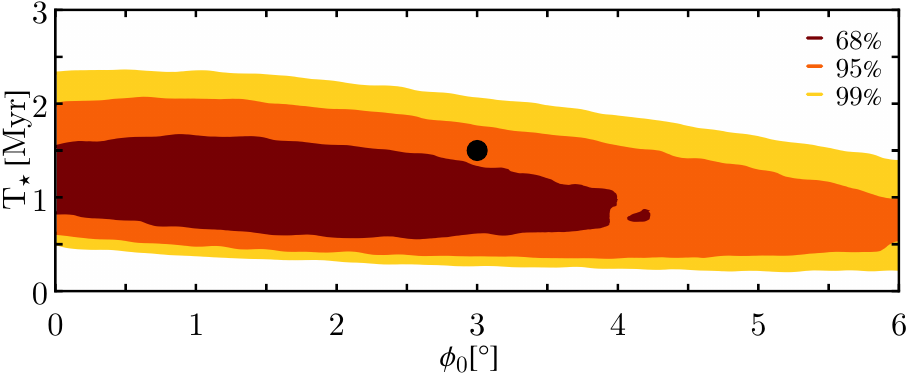}
\caption{Same as in Fig.~\ref{fig:CornerFiducial},
but for the two-population simulations from~\S\ref{sec:LessSimpleFiducial}.
See~\S\ref{sec:DiscDilution} for details.
}
\label{fig:CornerSgrA}
\end{figure}

We now turn our interest to a astrophysically more relevant
astrophysical system, namely the two-population (stars and \IMBHs\@)
Top-Heavy model from~\cite{Generozov+2020} (\S\ref{sec:LessSimpleFiducial}),
in which we inject discs of 7 test stars
mimicking the observed S-stars.
In Fig.~\ref{fig:CornerSgrA}, we illustrate our 
expected likelihood
as one varies respectively ${ (m_{\bullet} , \gamma_{\bullet}) }$
the individual mass and power-index of the \IMBH\ population
or ${ (\phi_{0} , T_{\star}) }$ the initial angular size of the disc
and their (common) stellar ages. 
We average over many realisations of the observations (\S\ref{sec:DiscDilution}).
We note that the reconstruction is not ideal
and suffers from some bias.
One should keep these limitations in mind
when applying \VRR\ neighbor separation to SgrA*'s
clockwise disc.

\section{Results for SgrA*}
\label{sec:Results}

We now apply our approach to the clockwise disc of S-stars (see~\S\ref{sec:StellarCusps}).
The realm of possible models for SgrA*'s background
is extremely large.
For the sake of simplicity,
we only consider
baths similar to the Top-Heavy model
from~\cite{Generozov+2020} (\S\ref{sec:LessSimpleFiducial}).
These clusters are therefore characterised
by two background populations (stars and \IMBHs\@)
each with their own individual mass and power-law index,
an initial angular separation for the disc, $\phi_{0}$ (see Eq.~\ref{def_vMF}),
and a common stellar age, ${ T_{\star} }$, for the S-stars.
For simplicity, we arbitrarily impose ${ \phi_{0} \!=\! 3^{\circ} }$,
and fix the age of the stars to ${ T_{\star} \!=\! 7.1\,\Myr }$ (see~\S\ref{sec:StellarCusps}).
We marginalise our likelihood over the observational 
uncertainties in the S-star's orbital parameters from~\cite{Gillessen+2017},
and over the initial orientation of the stellar orbits
within a narrow distribution around ${ \phi_{0} \!=\! 3^{\circ} }$,
following Eq.~\eqref{def_vMF}.

\begin{figure}
\centering
\includegraphics[width=0.45 \textwidth]{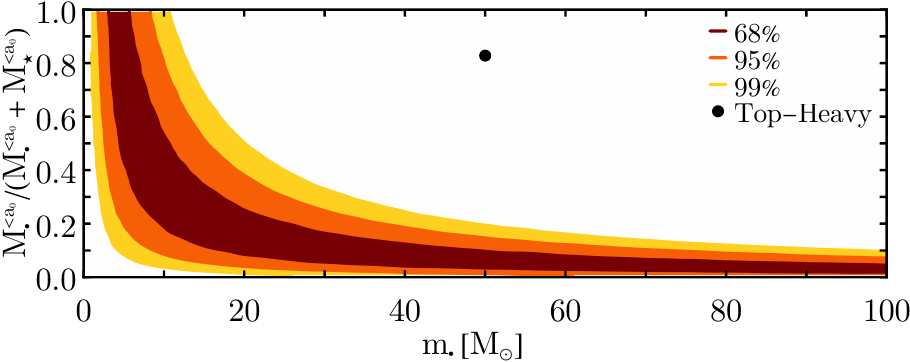}
\caption{Likelihood estimated from the S-stars' resolved
orbits. We assume that SgrA*'s background cluster follows the
two-population Top-Heavy model from~\S\ref{sec:LessSimpleFiducial},
and respectively vary
${ (m_{\bullet},M_{\bullet}(\!<\! a_{0})/(M_{\bullet} (\!<\! a_{0}) \!+\! M_{\star} (\!<\! a_{0}))) }$
with the total enclosed mass,
${ M_{\bullet} (\!<\! a_{0}) \!+\! M_{\star} (\!<\! a_{0}) }$,
and all the other parameters fixed.
The black dot corresponds to the Top-Heavy model~\citep{Generozov+2020}.
See~\S\ref{sec:Likelihood} for details.
}
\label{fig:CornerSstar_massMenc}
\end{figure}

In Fig.~\ref{fig:CornerSstar_massMenc},
we plot this likelihood, varying
the individual mass $m_{\bullet}$ as well as the
total IMBH mass fraction ${ M_{\bullet} (\!<\!a_{0})/(M_{\bullet}(\!<\! a_{0}) \!+\! M_{\star} (\!<\! a_{0})) }$,
keeping the mass enclosed within a fiducial scale radius $a_0$, 
${ M_{\bullet} (\!<\! a_{0}) \!+\! M_{\star} (\!<\! a_{0}) }$, fixed.
Interestingly, we find that the Top-Heavy model from~\cite{Generozov+2020} 
is incompatible with SgrA*'s observed stellar disc.
The survival of the disc through the VRR
dynamics requires indeed a quieter bath,
i.e.\ a smaller \IMBH\ fraction.

\begin{figure}
\centering
\includegraphics[width=0.45 \textwidth]{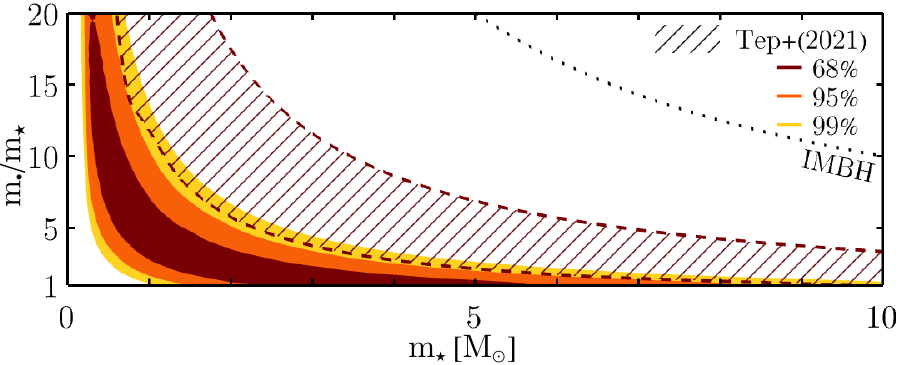}
\caption{Same as in Fig.~\ref{fig:CornerSstar_massMenc},
this time jointly varying
${ (m_{\star},m_{\bullet}/m_{\star}) }$
keeping all other parameters fixed. The dashed line
corresponds to ${ m_{\bullet} \!=\! 100 \Msun }$
above which the heavy objects are
usually considered as \IMBHs\@.
The hashed contours correspond to
the 68\% level lines constrained in~\protect\cite{Tep+2021}
by considering eccentricity relaxation.
}
\label{fig:CornerSstar_mratio}
\end{figure}

Finally, in Fig.~\ref{fig:CornerSstar_mratio},
we let the individual masses of the
background cluster, $m_{\star}$ and $m_{\bullet}$,
vary with the natural constraint ${ m_{\bullet} \!\geq\! m_{\star} }$,
while fixing the total enclosed masses
${M_{\star} (\!<\! a_{0})}$ and ${ M_{\bullet} (\!<\! a_{0}) }$.
In that figure, we find that the survival of SgrA*'s disc requires a drastically 
smaller individual mass for the background population of heavy particles.
Assuming that all the other parameters of the background
bath are fixed, the current presence of SgrA*'s disc seems
in tension with the presence of \IMBHs\@,
as illustrated by the
dotted line in that plot.

A similar parameter exploration has recently been presented
in the bottom right panel of fig.~{3} of~\cite{Tep+2021}.
There, the authors used the eccentricity relaxation of
the S-stars to constrain SgrA*'s stellar content, assuming
that the S-stars were initially born from an initial quasi-
circular disc. In Fig.~\ref{fig:CornerSstar_mratio},
we reproduce the bottom-right panel
of fig.~{3} of~\cite{Tep+2021}. It is very interesting to note
that both works, although they are using different dynamical
processes (scalar vs. vector resonant relaxation) offer similar
constraints on SgrA*'s background cluster. They jointly
constrain the likelihood of IMBHs orbiting SgrA*. Naturally,
one should keep in mind all the limitations of the present
exploration, in particular the small number of parameters
effectively probed. Future works will focus on exploring
larger parameter spaces by sampling from the full likelihood.

\section{Conclusions}
\label{sec:Conclusions}

We illustrated how a statistical characterisation of \VRR\
used jointly with the recent observation of SgrA*'s clockwise stellar disc
allows one to place constraints on the content of SgrA*'s underlying
stellar cluster, in particular \IMBHs\@. More precisely, to be compatible with the disc's being still present today,
\VRR\ fluctuations cannot be too large,
i.e.\ cannot be too efficient at driving the spontaneous dilution
of the disc.
Keeping in mind the small range of parameters explored,
we showed that the survival of SgrA*'s disc tends
to reduce the likelihood for heavy background particles,
like \IMBHs\@, to orbit around SgrA*.

Naturally, this paper is only a first step toward fully leveraging
\VRR\ to constrain SgrA*'s stellar content: a more thorough 
exploration of parameter space is a logical next step. We now conclude
with a few possible additional venues for future works.

The statistical description of neighbor separation by \VRR\ (\S\ref{sec:NeighSep})
relies extensively on the assumption of an isotropic background bath.
This is key to simplify the analytical calculations
However, this does not account for any possible anisotropic clustering in orientations,
such as \IMBHs\ discs~\citep{Szolgyen+2018}
and the associated resonant dynamical friction~\cite[see, e.g.\@,][]{Szolgyen+2021}.
Indeed, a rotating background stellar cluster is expected
to source (resonant) dynamical friction,
possibly leading to an efficient disruption of the disc~\citep{Levin2022}.
Going beyond isotropy,
e.g.\@, axisymmetric distributions,
will be the focus of future works.

We also assumed the limit of test particles for the disc stars,
i.e.\ we neglected the self-gravity among them.
Contributions from this additional coupling were already investigated
numerically in~\cite{Kocsis+2011}, see figs.~{6} and~{7} therein.
It showed that self-gravity increased the coherence of the disc,
hence delaying its dilution:
the more massive the disc,
the longer its survival.
Incorporating this additional effect
is no easy analytical undertaking.

We neglected the effect of SgrA*'s spin,
which via the Lense--Thirring precession~\citep[see, e.g.\@,][]{Merritt2013}
can also drive a spontaneous dilution of the disc~\citep{Levin+2003,Fragione+2022}.
We point out that these effects could still be somewhat small here
because the disc stars are quite far from SgrA*, ${ a_{\mathrm{disc}} \!\gtrsim\! 50 \, \mpc }$.
In addition, the appropriate timescale for dilution is the one of phase mixing,
i.e.\ the orbital planes of the disc stars must precess respectively
from one another. It would be worthwhile
to include this additional effect.

\VRR\ relies on a double orbit-average
over both the fast Keplerian motion and the in-plane precession.
As a result, the semi-major axes and eccentricities of each annuli
are supposed to be conserved. This amounts to neglecting
the possible contributions of scalar resonant relaxation
and non-resonant relaxation~\citep{Rauch+1996}.
Although slower, these additional processes
could still marginally impact the disc's dilution.

Finally, we expect that future observations,
such as ELT~\citep{Pott+2018,Davies+2018} and TMT~\citep{Do+2019},
will soon provide the community with a wealth
of resolved orbits around SgrA*, along with their stellar ages.
This additional statistics will prove paramount
to tighten the present constraints on SgrA*'s stellar content.
In particular, the same data will allow one
to constrain both the efficiency of eccentricity relaxation~\citep{Tep+2021}
as well as orientation relaxation (this work).
Ultimately, these two dynamical windows should offer
complementary constraints
on the statistics of the possible \IMBHs\
orbiting around SgrA*.

\subsection*{Data Distribution}
The code and the data underlying this article 
is available through reasonable request to the authors.

\section*{Acknowledgements}
JBF is partially supported by grant Segal ANR-19-CE31-0017
of the French Agence Nationale de la Recherche,
and by the Idex Sorbonne Universit\'e.
AZ was supported by NSF Grants PHY-1912578 and PHY-2207594.
JBF warmly thanks F.\ Leclercq for various suggestions.

\appendix

\section{Stellar cusps in SgrA*}
\label{sec:StellarCusps}

In this Appendix, we specify our convention for the description
of the background bath.
Following~\cite{Gillessen+2017}, we fix SgrA*'s mass to
${ \MBH \!=\! 4.28 \!\times\! 10^{6} \Msun }$.
We take SgrA*'s distance from the Earth to be ${ d \!=\! 8.178 \, \kpc}$~\citep{Abuter+2019}.
This is used to obtain the S-stars' semi-major axes
from table~{3} in~\cite{Gillessen+2017}.
Following fig.~{12} of~\cite{Gillessen+2017},
we assume that the disc is composed of a total of seven stars,
namely (S66, S67, S83, S87, S91, S96, S97).
Their ages are inferred from the main-sequence ages measured
in table~{2} of~\cite{Habibi+2017}, albeit for different S-stars.
In practice, we assume that the disc stars' ages are identical
and impose ${ T_{\star} \!=\! 7.1 \, \Myr }$,
the average stellar age of the S-stars with both resolved orbits
and constrained main-sequence stellar ages.
This large uncertainty on the stars' ages is one the main limitations
of the results from the main text.

A given bath is composed of different stellar
populations within which all the orbits share the same individual mass.
In pratice, we make two important assumptions:
(i) each population follows a power-law distributin in semi-major axes;
(ii) each population follows a thermal distribution in eccentricities.
As a consequence, a given population is characterised by three numbers,
namely (i) $\gamma$, the slope of the power-law profile in semi-major axis;
(ii) $m$, the individual mass of the particles;
(iii) ${ M (\!<\! a_{0}) }$, the total mass physically enclosed
within a radius $a_{0}$, a radius of reference.

More precisely, for a given population the number of stars
per unit semi-major axis, $a$,
is given by
\begin{equation}
N (a) = (3 \!-\! \gamma) \, \frac{N_{0}}{a_{0}} \, \bigg( \frac{a}{a_{0}} \bigg)^{2 - \gamma} ,
\label{PDF_N}
\end{equation}
where we introduced ${ N_{0} \!=\! g(\gamma) \, N(\!<\! a_{0}) }$ with
\begin{equation}
g(\gamma) = 2^{- \gamma} \, \sqrt{\pi} \, \Gamma (1 \!+\! \gamma) / \Gamma (\gamma \!-\! \half) ,
\label{def_g}
\end{equation}
with ${ \Gamma (x) }$ the gamma function
and ${ N (\!<\! a_{0}) \!=\! M(\!<\! a_{0}) / m }$ the number of stars
physically enclosed within the radius $a_{0}$.
We assume that the distribution of eccentricities is thermal,
neglecting effects associated with the SgrA*'s loss cone.
Eccentricities follow then the \PDF\
\begin{equation}
f (e) = 2 e ,
\label{PDF_thermal}
\end{equation}
which satisfies the normalisation ${ \!\int\! \rd e f (e) \!=\! 1 }$.
Following the convention from~\S\ref{sec:VRR},
the \DF\ of a given background population is therefore
\begin{equation}
n (\bKp)  = \tfrac{1}{4 \pi} \, \deltaD (m^{\prime} \!-\! m) \, N (a^{\prime}) \, f (e^{\prime}) ,
\label{full_nK}
\end{equation}
recalling that the background populations are assumed to be isotropic on average.

Assuming that the background populations follow power law distributions
greatly simplifies the computation of the coherence time,
see~\S{I} in~\cite{Fouvry+2019}. In practice, following the notations therein,
we pre-computed the dimensionless function ${ f_{\Gamma^{2}} (e) }$
for ${ 0 \!\leq\! \gamma \!\leq\! 3 }$ and ${ 0 \!\leq\! e \!\leq\! 0.99 }$,
on a ${ 300\!\times\! 100 }$ grid,
to be interpolated afterwards.
This allows for efficient evaluations of the coherence time
of the background bath, ${ \Tc (\bK) }$,
as defined in Eq.~\eqref{def_Tc}.

\section{\VRR\ neighbor separation}
\label{sec:NeighSep}

\begin{figure}
\centering
\includegraphics[width=0.45 \textwidth]{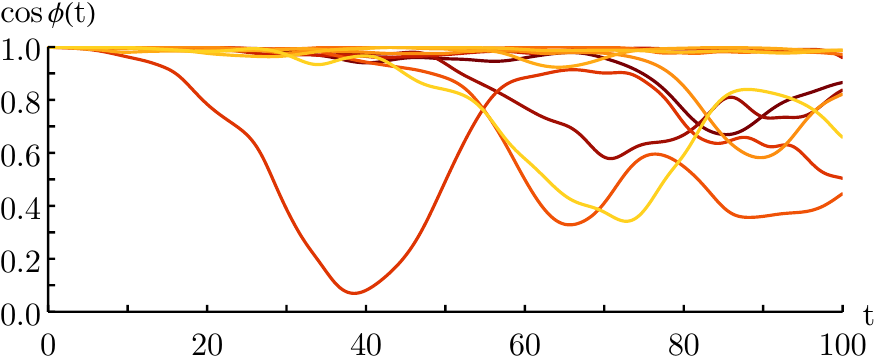}
\caption{Typical random walks of pairwise angular
separations, $\cos \phi (t)$, as observed in the simple setup
from~\S\ref{sec:SimpleFiducial}.
}
\label{fig:PairsFiducial}
\end{figure}

We are interested in the stochastic evolution of the angle
${ \phi (t) }$ between two test stars' orbits of parameters ${ (\bK_{1} , \bK_{2}) }$
embedded within a background bath characterised
by ${ n(\bK) }$. Figure~\ref{fig:PairsFiducial} illustrates typical
random walks undergone by ${ \cos \phi (t) }$ for the simple bath
from~\S\ref{sec:SimpleFiducial}.  

\subsection{Analytical prediction}
\label{sec:AnalyticalPrediction}

Following~{G20}, the dilution of the two test stars
can be described via the moments
of the correlation function of the stars' respective orientations,
namely
\begin{equation}
t \mapsto \big\langle P_{\ell_{\alpha}} (\cos \phi (t)) \big\rangle ,
\label{def_moments}
\end{equation}
with $P_{\ell_{\alpha}}$ the Legendre polynomials
and ${ \langle \, \cdot \, \rangle }$ standing for an ensemble
average over realisations of the bath and initial conditions
of the two test stars. Following Eq.~{(21)} of~G20,
these moments evolve according to\footnote{Equation~{(21)} of~G20
is missing a ${1/(4\pi)}$ in its rhs.}
\begin{equation}
\big\langle P_{\ell_{\alpha}} (\cos \phi (t)) \big\rangle = D_{\ell_{\alpha}} \, C_{\ell_{\alpha}}^{\bO} (t) \, C_{\ell_{\alpha}}^{\bD} (t) .
\label{generic_rate}
\end{equation}

In Eq.~\eqref{generic_rate}, ${ D_{\ell_{\alpha}} \!=\! \langle P_{\ell_{\alpha}} (\cos \phi (0)) \rangle }$ describes
the statistics of the angular separation at ${ t \!=\! 0 }$.
Following Eq.~{(D9)} of~G20,
the function ${ C_{\ell_{\alpha}}^{\bO} (t) }$ describes the separation
driven by the differences in the two particles orbital parameters,
i.e.\ driven by ${ \bK_{1} \!\neq\! \bK_{2} }$.
It reads
\begin{equation}
C_{\ell_{\alpha}}^{\bO} (t) = \exp \!\bigg[\! - \half A_{\ell_{\alpha}} \sum_{\ell} B_{\ell} \, I_{\ell}^{-} (\bK_{1} , \bK_{2} , t) \bigg] ,
\label{def_CO}
\end{equation}
Similarly, following Eq.~{(D10)} of~{G20}\footnote{Equation~{(D10)} of G20 is missing a minus sign in its rhs.},
the function ${ C_{\ell_{\alpha}}^{\bD} (t) }$ captures the separation sourced
by the differences in the test particles' initial orientations.
It reads
\begin{equation}
C_{\ell_{\alpha}}^{\bD} (t) \!=\! \exp \!\bigg[\! - \!\sum_{\ell , \ell_{\gamma}}\! \frac{D_{\ell_{\alpha}} \!\!-\! D_{\ell_{\gamma}}}{(2 \ell_{\alpha} \!+\! 1) D_{\ell_{\alpha}}} \big( E_{\ell_{\alpha} \ell \ell_{\gamma}}^{L} \big)^{2} I_{\ell}^{+} (\bK_{1} , \bK_{2} , t) \bigg] .
\label{def_CD}
\end{equation}

In Eqs.~\eqref{def_CO} and~\eqref{def_CD},
we defined the integrals
\begin{align}
I_{\ell}^{-} (\bK_{1} , \bK_{2} , t) {} & = \!\! \int \!\! \rd \bK \, n (\bK) \, \big( \mJ_{\ell} [\bK_{1} , \bK] \!-\! \mJ_{\ell} [\bK_{2} , \bK] \big)^{2}
\nonumber
\\
{} & \times \frac{2 \Tc^{2} (\bK)}{A_{\ell}} \chi \big[ \sqrt{A_{\ell}/2} \, (t / \Tc (\bK)) \big] ,
\nonumber
\\
I_{\ell}^{+} (\bK_{1} , \bK_{2} , t) {} & = \!\! \int \!\! \rd \bK \, n(\bK) \, \mJ_{\ell} [\bK_{1} , \bK_{2}] \, \mJ_{\ell} [\bK_{2} , \bK] 
\nonumber
\\
{} & \times \frac{2 \Tc^{2} (\bK)}{A_{\ell}} \chi \big[ \sqrt{A_{\ell} / 2} \, (t / \Tc (\bK)) \big]  .
\label{def_I_minus_plus}
\end{align}
We also introduced the coefficients
\begin{equation}
A_{\ell} = \ell (\ell \!+\! 1) ; 
\quad
B_{\ell} = \tfrac{1}{8 \pi} \, \ell (\ell + 1) (2 \ell + 1) ,
\label{def_Al_Bl}
\end{equation}
along with the bath's coherence time
\begin{equation}
\frac{1}{\Tc^{2} (\bK)} = \sum_{\ell} B_{\ell} \!\! \int \!\! \rd \bKp \, n (\bKp) \, \mJ_{\ell}^{2} [\bK , \bKp ] ,
\label{def_Tc}
\end{equation}
where the coupling coefficients, ${ \mJ_{\ell} [\bK , \bKp] }$,
are given in~\S{A} of~{G20}.
In practice, these coefficients are pre-computed using
the same approach as in~\cite{Magnan+2022}.
Equation~\eqref{def_CD} also involves the isotropic Elsasser coefficients,
${ E_{\ell_{\alpha} \ell \ell_{\gamma}}^{L} }$,
as spelled out in~\S{B} of~\cite{Fouvry+2019}.
Finally, we introduced the dimensionless function
\begin{equation}
\chi (\tau) = \!\! \int_{0}^{\tau} \!\!\!\! \rd \tau_{1} \!\! \int_{0}^{\tau} \!\!\!\! \rd \tau_{2} \, \re^{- (\tau_{1} - \tau_{2})^{2}} = \re^{- \tau^{2}} \!-\! 1 \!+\! \sqrt{\pi} \, \tau \, \erf (\tau) ,
\label{def_chi}
\end{equation}
which captures the transition between
the ballistic and diffusive regimes~\citep{Fouvry+2019}.

Unfortunately, as argued in~\S{4} of~{G20},
the analytical prediction from Eq.~\eqref{generic_rate}
does not provide a good match to the late-time behavior
of the dilution of test stars that start with very similar
initial orientations. To alleviate this difficulty,
we follow~{G20} and rely on a ``piecewise'' prediction.

\subsection{Piecewise prediction}
\label{sec:Piecewise}

Regardless of the initial similarity between the test particles,
Eq.~\eqref{generic_rate} works well on short timescales,
i.e.\ for ${ t \!\lesssim\! \Tc (\bK) }$. As such, we give ourselves
the timelapse
\begin{equation}
\Delta t = \Min [\Tc (\bK_{1}) , \Tc (\bK_{2})] ,
\label{def_Delta}
\end{equation}
and set out to construct the sequence of angular separations
\begin{equation}
\cos \phi_{0} \to ... \to \cos \phi_{n} \!=\! \cos \phi(t \!=\! n \Delta t) .
\label{def_sequence}
\end{equation}
This prediction is Markovian in the sense that we treat
the transition ${ \cos \phi_{i} \!\to\! \cos \phi_{i + 1} }$
as independent of the past history of the test particles.

Following Eq.~{(G1)} of~{G20}, we can write
\begin{equation}
\big\langle P_{\ell} (\cos \phi_{i+1}) \big\rangle = \!\! \int \!\! \rd (\cos \phi_{i}) \, \rho_{i} (\cos \phi_{i}) \, \big\langle P_{\ell} (\cos \phi_{i+1}) \,\big|\, \cos \phi_i \big\rangle .
\label{moment_Pl_iteration}
\end{equation}
In that expression, ${ \langle P_{\ell} (\cos \phi_{i+1}) \,|\, \cos \phi_i \rangle }$
follows from Eq.~\eqref{generic_rate} evaluated after a time ${ t \!=\! \Delta t }$
and assuming that the test particles are initially separated by the fixed angle $\phi_i$,
i.e.\ ${ D_{\ell} \!=\! P_{\ell} (\cos \phi_{i}) }$.
Equation~\eqref{moment_Pl_iteration} also involves ${ \rho_{i} (\cos \phi_{i}) }$,
the \PDF\ of ${ \cos \phi_{i} }$, which is unknown.
Our goal is therefore to average over it.

\subsection{Perturbative expansion}
\label{sec:Perturbative}

We now make progress
by performing a perturbative expansion of Eq.~\eqref{generic_rate}
for small angles.\footnote{Here, we improve upon~{G20}
and perform a second-order expansion. Though,
the crux of the calculations remains identical.}
For ${ \cos \phi_{0} \!\to\! 1 }$, the Legendre polynomials
obey the second-order expansion
\begin{equation}
P_{\ell} (\cos \phi_{0}) \simeq 1 \!-\! \half A_{\ell} (1 \!-\! \cos \phi_{0}) \!+\! \tfrac{1}{16} C_{\ell} (1 \!-\! \cos \phi_{0})^{2} ,
\label{DL_Legendre}
\end{equation}
with ${ C_{\ell} \!=\! A_{\ell} (A_{\ell} \!-\! 2) }$.
Let us now expand Eq.~\eqref{def_CD} at second order
in ${ 1 \!-\! \cos \phi_0 }$. We write
\begin{align}
\frac{D_{\ell_{\alpha}} \!\!-\! D_{\ell_{\gamma}}}{(2 \ell_{\alpha} \!+\! 1) D_{\ell_{\alpha}}} \simeq {} & \half \big( 1 \!-\! \cos \phi_{0} \big) \, \frac{A_{\ell_{\gamma}} \!\!-\! A_{\ell_{\alpha}}}{2 \ell_{\alpha} \!+\! 1}
\label{expansion_sum}
\\
+ {} & \tfrac{1}{16} \big( 1 \!-\! \cos \phi_{0} \big)^{2} \frac{C_{\ell_{\alpha}} \!\!-\! C_{\ell_{\gamma}} \!+\! 4 A_{\ell_{\alpha}} (A_{\ell_{\gamma}} \!\!-\! A_{\ell_{\alpha}})}{2 \ell_{\alpha} \!+\! 1} ,
\nonumber
\end{align}
where we used ${ D_{\ell} \!=\! P_{\ell} (\cos \phi_{0}) }$.
Pursuing the calculation, we can perform the sum over $\ell_{\gamma}$
in Eq.~\eqref{def_CD} to get
\begin{align}
\sum_{\ell_{\gamma}} {} & \!\frac{D_{\ell_{\alpha}} \!\!-\! D_{\ell_{\gamma}}}{(2 \ell_{\alpha} \!+\! 1) D_{\ell_{\alpha}}} \big( E_{\ell_{\alpha} \ell \ell_{\gamma}}^{L} \big)^{2} \simeq \half \big( 1 \!-\! \cos \phi_{0} \big) \, A_{\ell_{\alpha}} B_{\ell} (A_{\ell} \!-\! 2)
\nonumber
\\
{} & \quad + \tfrac{1}{16} \big( 1 \!-\! \cos \phi_{0} \big)^{2} A_{\ell_{\alpha}} B_{\ell} (A_{\ell} \!-\! 2) (A_{\ell_{\alpha}} \!\!-\! A_{\ell} \!+\! 6) .
\label{expansion_sum_sum}
\end{align}
To get this expression, we used some contraction rules of the isotropic
Elsasser coefficients, namely Eqs.~{(B3)} and~{(B4)} in~{G20},
along with
\begin{align}
\sum_{\ell_{\gamma}} \tfrac{1}{2 \ell_{\alpha} + 1} {} & \big( E_{\ell_{\alpha} \ell \ell_{\gamma}}^{L} \big)^{2} C_{\ell_{\gamma}} = A_{\ell_{\alpha}} B_{\ell}
\nonumber
\\
{} & \times \big[ 12 \!-\! 6 A_{\ell} \!-\! 6 A_{\ell_{\alpha}} \!\!+\! 3 A_{\ell} A_{\ell_{\alpha}} \!+\! C_{\ell} \!+\! C_{\ell_{\alpha}} \big] .
\label{contract_Elsasser}
\end{align}

At second-order in ${ (1 \!-\! \cos \phi_{0}) }$,
Eq.~\eqref{generic_rate} becomes
\begin{equation}
\big\langle P_{\ell_{\alpha}} (\cos \phi (t)) \,\big|\, \cos \phi_{0} \big\rangle \simeq P_{\ell_{\alpha}} (\cos \phi_{0}) \, C_{\ell_{\alpha}}^{\bO} (t) \, C_{\ell_{\alpha}}^{\bD} (t) ,
\label{rewrite_expansion_P}
\end{equation}
with
\begin{equation}
C_{\ell_{\alpha}}^{\bO} (t) \simeq \exp \!\big[\! - \half A_{\ell_{\alpha}} \, \Psi^{-} (\bK_{1} , \bK_{2} , t ) \big] ,
\label{expansion_CO}
\end{equation}
and
\begin{align}
{} & C_{\ell_{\alpha}}^{\bD} (t) \simeq \exp \!\big[\! - \half A_{\ell_{\alpha}} (1 \!-\! \cos \phi_{0}) \, \Psi^{+} \!(\bK_{1} , \bK_{2} , t)
\label{expansion_CD}
\\
- {} & \;  \tfrac{1}{16} A_{\ell_{\alpha}} \!(1 \!-\! \cos \phi_{0})^{2} \big( A_{\ell_{\alpha}} \!\Psi^{+} \!(\bK_{1} , \bK_{2} , t) \!-\! \Psi^{++} \!(\bK_{1} , \bK_{2} , t) \big) \big] .
\nonumber
\end{align}
In these expressions, we introduced
\begin{align}
\Psi^{-} (\bK_{1} , \bK_{2} , t)  = {} & \sum_{\ell} B_{\ell} \, I_{\ell}^{-} (\bK_{1} , \bK_{2} , t) ,
\label{def_Psi}
\\
\Psi^{+} (\bK_{1} , \bK_{2} , t) = {} & \sum_{\ell} B_{\ell} (A_{\ell} \!-\! 2) \, I^{+}_{\ell} (\bK_{1} , \bK_{2} , t) ,
\nonumber
\\
\Psi^{++} (\bK_{1} , \bK_{2} , t) = {} & \sum_{\ell} B_{\ell} (A_{\ell} \!-\! 2) (A_{\ell} \!-\! 6) \, I^{+}_{\ell} (\bK_{1} , \bK_{2} , t)
\nonumber
\end{align}
In practice, the integrals over ${ (a,e) }$ appearing in Eq.~\eqref{def_Psi}
are computed using the standard midpoint rule
with ${ K_{a} \!=\! 100 }$ nodes sampled uniformly in logarithm in the domain
${ \Max[a_{1}, a_{2}] \!\times\! 10^{-3} \!\leq\! a \!\leq\! \Min[a_{1},a_{2}] \!\times\! 10^{3} }$,
and using ${ K_{e} \!=\! 20 }$ nodes uniformly
within the domain in $e$ of the considered system.

\subsection{Propagating the moments}
\label{sec:Propagating}

We continue the calculation by expanding
explicitly the rhs of Eq.~\eqref{rewrite_expansion_P} at second order
in ${ (1 \!-\! \cos \phi_{0}) }$. We have
\begin{equation}
\big\langle P_{\ell_{\alpha}} \!(\cos \phi (t)) \,\big|\, \cos \phi_{0} \big\rangle \simeq
a_{\ell_{\alpha}} \!+\! b_{\ell_{\alpha}} P_{1} (\cos \phi_{0}) \!+\! c_{\ell_{\alpha}} P_{2} (\cos \phi_{0}) .
\label{DL_expansion_P}
\end{equation}
In that expressions, the coefficients ${ (a_{\ell} , b_{\ell} , c_{\ell}) }$ read
\begin{align}
a_{\ell} {} & = u_{\ell} + \tfrac{1}{12} A_{\ell} u_{\ell} \big\{ v_{\ell} - 6 \Psi^{+} \!- 8 \big\} ,
\nonumber
\\
b_{\ell} {} & = - \tfrac{1}{8} A_{\ell} u_{\ell} \big\{ v_{\ell} - 4 \Psi^{+} \!- 6 \big\} ,
\nonumber
\\
c_{\ell} {} & = \tfrac{1}{24} A_{\ell} u_{\ell} \big\{ v_{\ell} - 2 \big\} ,
\label{exp_abc}
\end{align}
where we introduced
\begin{equation}
u_{\ell} = \re^{- \tfrac{1}{2} A_{\ell} \Psi^{-}} ;
\quad
v_{\ell} = \Psi^{++} \!+ A_{\ell} (1 \!+\! \Psi^{+}) (1 \!+\! 2 \Psi^{+}) ,
\label{def_uv}
\end{equation}
and used the shortened notation ${ \Psi^{-} \!=\! \Psi^{-} (\bK_{1} , \bK_{2} , t) }$,
and similarly for $\Psi^{+}$ and $\Psi^{++}$.

We now have everything at our disposal to predict jointly
both moments ${ \langle P_{1} (\cos \phi_{i}) \rangle }$
and ${ \langle P_{2} (\cos \phi_{i}) \rangle }$. Let us introduce the vector and matrices
\begin{equation}
\bP_{i} \!=\!
\begin{pmatrix}
\langle P_{1} (\cos \phi_{i}) \rangle
\\
\langle P_{2} (\cos \phi_{i}) \rangle
\end{pmatrix} ;
\;
\bA \!=\!
\begin{pmatrix}
a_{1}
\\
a_{2}
\end{pmatrix} ;
\;
\bM \!=\!
\begin{pmatrix}
b_{1} & c_{1}
\\
b_{2} & c_{2}
\end{pmatrix} ,
\label{def_P_A_M}
\end{equation}
so that Eq.~\eqref{DL_expansion_P},
evaluated at time ${ t \!=\! \Delta t }$,
reads
\begin{equation}
\bP_{i + 1} = \bA \!+ \bM \, \bP_{i} .
\label{vec_iteration}
\end{equation}
The fixed point of this iteration
is ${ \bR \!=\! [\bI \!-\! \bM]^{-1} \bA }$,
and can be used to construct the generic solution
\begin{equation}
\bP_{i} = \bR + \bM^{i} \, \big[ \bP_{0} \!-\! \bR \big] .
\label{sol_iteration}
\end{equation}
To obtain a ``continuous'' prediction from this discrete solution,
we diagonalise $\bM$ via ${ \bM \!=\! \bU \, \Diag[\lambda_{1} , \lambda_{2}] \, \bU^{-1} }$,
which gives ${ \bM^{i} \!=\! \bU \, \Diag[\lambda_{1}^{i} , \lambda_{2}^{i}] \, \bU^{-1} }$.
To obtain a continuous prediction,
we finally make the replacement ${ i \!\to\! t / \Delta t }$,
assuming that the two eigenvalues ${ (\lambda_{1} , \lambda_{2}) }$
are either both positive or complex conjugate\footnote{This is not guaranteed a priori by the present expansion.
In practice, for all the background bath and pairs of test particles considered,
this was, fortunately, always the case.}.

\begin{figure}
\centering
\includegraphics[width=0.45 \textwidth]{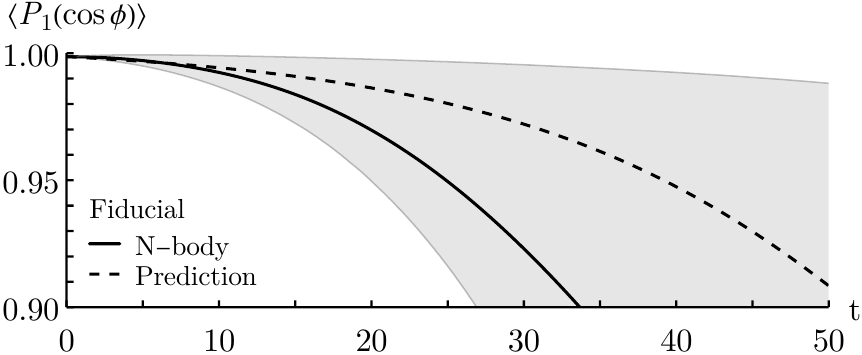}\\
\includegraphics[width=0.45 \textwidth]{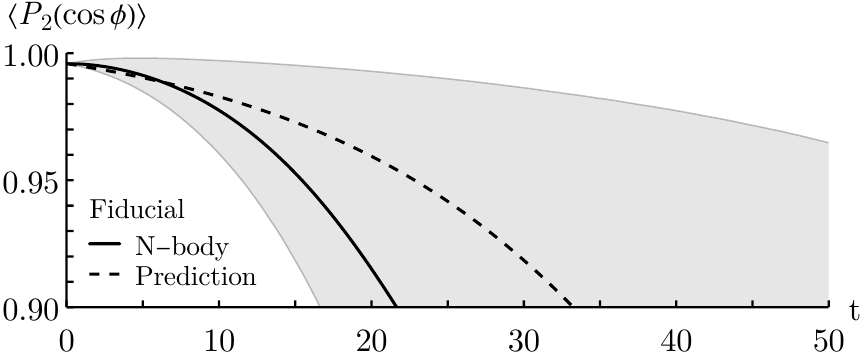}
\caption{Time evolution of the two first moments
${ \langle P_{1} (\cos \phi) \rangle }$ (top)
and ${ \langle P_{2} (\cos \phi) \rangle }$ (bottom)
for the fiducial bath from~\S\ref{sec:SimpleFiducial}.
The solid line is measured in numerical simulations,
with the gray region highlighting the 16\% and 84\% levels
among the available realisations.
The dashed line is the piecewise prediction from~\S\ref{sec:Propagating}.
}
\label{fig:P1P2Fiducial}
\end{figure}

In Fig.~\ref{fig:P1P2Fiducial}, we compare the initial time-evolution
of the first two moments, ${ \langle P_{1} (\cos \phi (t)) \rangle }$
and ${ \langle P_{2} (\cos \phi (t)) \rangle}$,
between the present kinetic modelling and fiducial numerical simulations (\S\ref{sec:SimpleFiducial}).
While the match is not ideal, we emphasise that the prediction
still succeeds at predicting the slow initial onset of separation,
i.e.\ the plateau that the naive application of the analytical
formula from Eq.~\eqref{generic_rate} cannot recover.

\begin{figure}
\centering
\includegraphics[width=0.45 \textwidth]{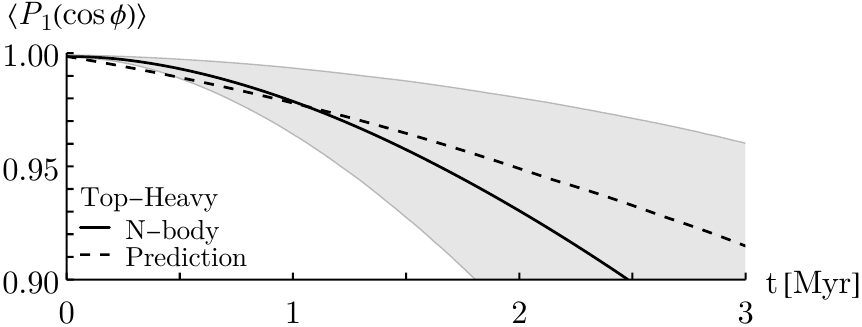}\\
\includegraphics[width=0.45 \textwidth]{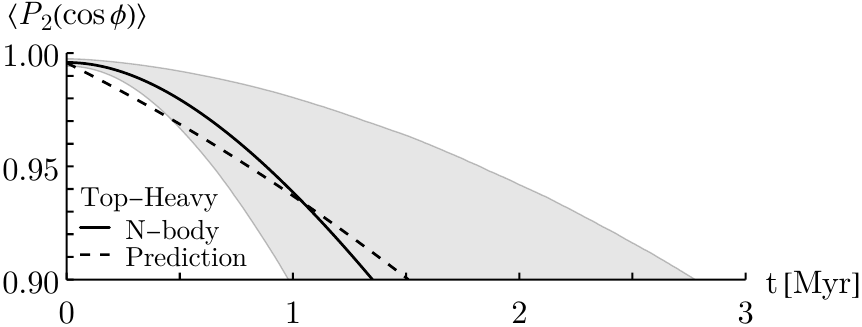}
\caption{Same as in Fig.~\ref{fig:P1P2Fiducial},
but for the two-population bath from~\S\ref{sec:LessSimpleFiducial}.
Here, ${ \langle P_{\ell} (\cos \phi) \rangle }$
is averaged over all the pairs of stars in discs composed of 7 stars each.
See the text for details.
}
\label{fig:P1P2SgrA}
\end{figure}

In Fig.~\ref{fig:P1P2SgrA},
we illustrate the same measurement
for the more realistic Top-Heavy model (\S\ref{sec:LessSimpleFiducial}).
In that case, every disc is composed of 7 stars
and sustains therefore 21 independent pairs of stars.
In Fig.~\eqref{fig:P1P2SgrA},
${ \langle P_{\ell} (\cos \phi) \rangle }$ stands therefore
for the average over all these pairs of stars.
To compute the kinetic prediction, following Eq.~\eqref{def_vMF},
we drew 100 independent initial conditions
for 100 discs, following Eq.~\eqref{def_vMF}.
These
are then used as initial conditions to compute
the prediction from~\S\ref{sec:Propagating}.
Finally, we average the Legendre moments over this large sample of pairs.
Reassuringly, Figs.~\ref{fig:P1P2Fiducial} and~\ref{fig:P1P2SgrA}
exhibit similar trends.

\subsection{Log-normal ansatz}
\label{sec:Lognormal}

We now have at our disposal a description for the initial time-evolution
of the first two moments
${ \langle P_{1} (\cos \phi (t)) \rangle }$
and ${ \langle P_{2} (\cos \phi (t)) \rangle }$.
Inspired by~\S{I1} in~{G20}, we now impose the ansatz
that the random variable ${ \phi (t) }$ follows a log-normal \PDF\
at all time. This is a strong assumption.
The \PDF\ reads
\begin{equation}
P (\phi \,|\, \mu , \sigma) = \frac{1}{\phi \, \sigma \sqrt{2 \pi}} \exp \!\big[\! - (\ln \phi - \mu)^{2} / (2 \sigma^{2})  \big] ,
\label{lognormal_PDF}
\end{equation}
for some parameters $\mu$, $\sigma$.

Unfortunately, our predictions for ${ (\langle P_{1} \rangle , \langle P_{2} \rangle ) }$
are not guaranteed to be realisable,
i.e.\ there might not exist a well-defined positive \PDF\ complying
with these provided moments.
Therefore,
to estimate ${ (\mu,\sigma) }$ from ${ (\langle P_{1} \rangle , \langle P_{2} \rangle) }$,
we proceed by minimising the distance
${ d \!=\! (\langle P_{1} \rangle \!-\! P_{1} [\mu ,\sigma])^{2} \!+\! (\langle P_{2} \rangle \!-\! P_{2} [\mu , \sigma] )^{2} }$ over ${ (\mu, \sigma) }$. In that expression,
the moments ${ P_{\ell} [\mu , \sigma] }$
are efficiently approximated following eq.~{(1.3)} of~\cite{Asmussen+2017}.
More precisely, we use
\begin{align}
\langle \cos (\ell \phi) \rangle {} & = \RePart \bigg[ \!\! \int_{0}^{+ \infty} \!\!\!\! \rd \phi \, \re^{\ri \ell \phi} \, P (\phi \,|\, \mu , \sigma ) \bigg]
\nonumber
\\
{} & \simeq \RePart \big[ \re^{- w (2 + w)/(2 \sigma^{2})} / \sqrt{1 \!+\! w} \big] ,
\label{approx_moments_lognormal}
\end{align}
with ${ w \!=\! W [- \ri \ell \sigma^{2} \re^{\mu}] }$,
${ W [z] }$ the Lambert function,
and the Legendre moments naturally follow.
This same expression can also be used to compute
the gradients ${ \p P_{\ell} [\mu,\sigma] / \p (\mu , \sigma) }$,
which are used by the optimiser.
In practice, the optimisation is performed using the LBFGS algorithm
from \texttt{Optim.jl}~\citep{Mogensen+2018}.

\begin{figure}
\centering
\includegraphics[width=0.45 \textwidth]{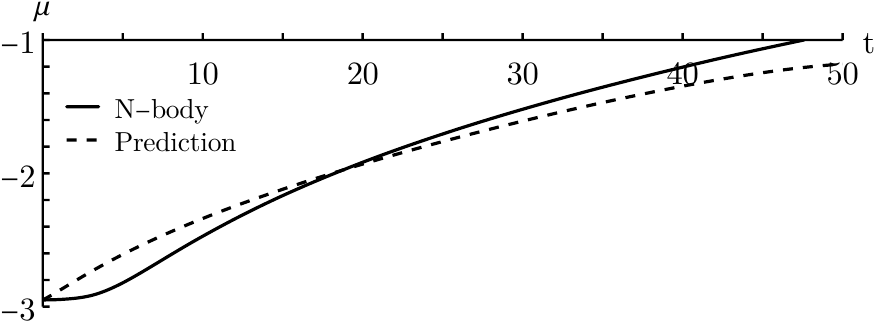}\\
\includegraphics[width=0.45 \textwidth]{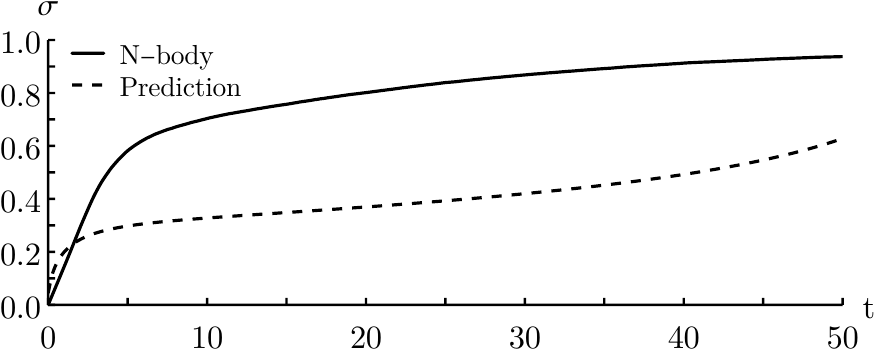}
\caption{Time evolution of the log-normal parameters
${ \mu }$ (top)
and ${ \sigma }$ (bottom)
for the fiducial simulations from~\S\ref{sec:SimpleFiducial}.
The solid line is fitted from the numerical simulations,
while the dashed line is the prediction from~\S\ref{sec:Lognormal}.
}
\label{fig:musigmaFiducial}
\end{figure}

In Fig.~\ref{fig:musigmaFiducial}, we compare the time-evolution
of the two lognormal parameters, ${ (\mu,\sigma) }$,
as obtained from the kinetic prediction
and directly fitted from the numerical simulations.
The match for $\mu$ is quite good,
while the \VRR\ formalism somewhat fails at predicting
${ \sigma }$.

\begin{figure}
\centering
\includegraphics[width=0.45 \textwidth]{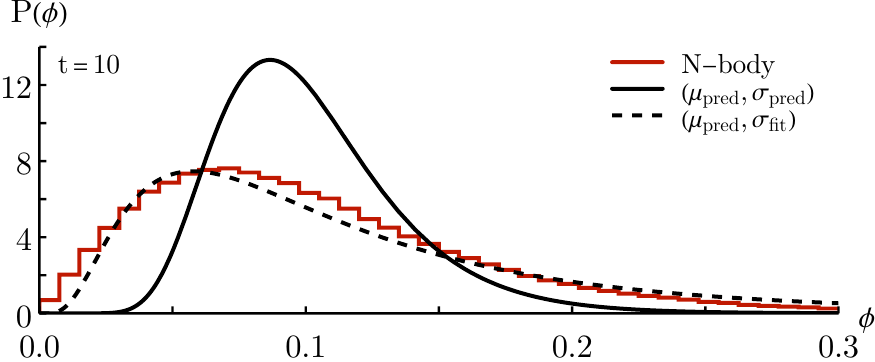}\\
\includegraphics[width=0.45 \textwidth]{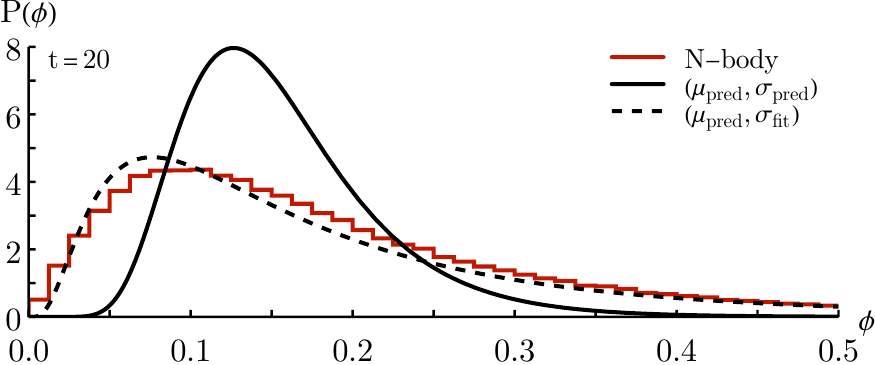}
\caption{Illustration of the \PDF\ of the angle ${ \phi (t) }$
in the fiducial simulations from~\S\ref{sec:SimpleFiducial}
for ${ t \!=\! 10 }$ (top) and ${ t \!=\! 20 }$ (bottom).
The full lines correspond to the prediction using both ${ (\mu,\sigma) }$
predicted in~\S\ref{sec:Lognormal},
while the dashed lines only use the predicted ${ \mu }$
and subsequently fit $\sigma$ from the empirical histograms.
}
\label{fig:HistogramFiducial}
\end{figure}

Finally, in Fig.~\ref{fig:HistogramFiducial},
we illustrate the \PDF\ of the angles ${ \phi }$
at various times, and compare it
with the predicted log-normal ansatz.
In that figure, we note that using the predicted $\mu$
and fitting $\sigma$ from the \PDF\ offers a very satisfactory match.
This motivates the marginalisation over $\sigma$
when performing likelihood estimations (\S\ref{sec:MarginalSigma}).

\subsection{Stitching}
\label{sec:Stitching}

Unfortunately, because it relies on a perturbative expansion
for small angles (\S\ref{sec:Perturbative}),
the piecewise prediction from Eq.~\eqref{sol_iteration}
cannot be used for arbitrarily large angular separation,
see, e.g.\@, fig.~{6} in~{G20}.
To circumvent this issue, we use Eq.~\eqref{sol_iteration} only
if the following conditions are all met:
(i) ${ \langle \cos \phi \rangle \!\geq\! \langle \cos \phicut \rangle \!=\! 0.8 }$,
i.e.\ the two test particles are close neighbors;
(ii) both ${ \p \langle P_{1} (\cos \phi (t)) \rangle / \p t }$
and ${ \p \langle P_{2} (\cos \phi (t)) \rangle / \p t }$ are negative,
i.e.\ the correlations decay.
In any other situations, we revert back to the straightforward application
of the analytical result from Eq.~\eqref{generic_rate}.
We now detail how this ``stitching'' between the piecewise and analytical
predictions can be made in practice.

Equation~\eqref{def_CO} is straightforward to evaluate
for ${ \ell_{\alpha} \!=\! 1, 2 }$.
To evaluate Eq.~\eqref{def_CD},
we first estimate the values of ${ D_{\ell} \!=\! \langle P_{\ell} (\cos \phi) \rangle }$
using Eq.~\eqref{approx_moments_lognormal}.
We then use the exclusion rules of the isotropic Elsasser coefficients,
${ E_{\ell_{\alpha} \ell \ell_{\gamma}}^{L} }$, see Eq.~{(B2)} in~{G20},
to perform the sum over $\ell_{\gamma}$ in Eq.~\eqref{def_CD}.
More precisely, for the particular cases ${ \ell_{\alpha} \!=\! 1 }$
(resp.\ ${ \ell_{\alpha} \!=\! 2 }$),
the range of this sum reduces to ${ \ell_{\gamma} \!=\! \ell }$
(resp.\ ${ \ell_{\gamma} \!=\! \ell \!-\! 1, \ell \!+\! 1 }$).
We also rely on the simple relations
\begin{align}
\big( E_{1 \ell \ell}^{L} \big)^{2} {} & = 6 B_{\ell} ,
\nonumber
\\
\big( E_{2,\ell,\ell - 1}^{L} \big)^{2} {} & = 15 B_{\ell} - \tfrac{45}{8 \pi} A_{\ell} ,
\nonumber
\\
\big( E_{2,\ell,\ell+1}^{L} \big)^{2} {} & = 15 B_{\ell} + \tfrac{45}{8 \pi} A_{\ell} .
\label{simple_EL}
\end{align}
We ultimately obtain the explicit expressions
\begin{align}
C_{1}^{\bD} (t) {} & \!=\! \exp \!\bigg[\! - \sum_{\ell} B_{\ell} \frac{2 (D_{1} \!-\! D_{\ell})}{D_{1}} I_{\ell}^{+} (\bK_{1} , \bK_{2} , t) \bigg] ,
\nonumber
\\
C_{2}^{\bD} (t) {} & \!=\! \exp \!\bigg[\! - \!\sum_{\ell}\! \frac{6 B_{\ell}}{D_{2}} \big( D_{2} \!-\! D_{\ell - 1} \tfrac{\ell - 1}{2 \ell + 1} \!-\! D_{\ell + 1} \tfrac{\ell + 2}{2 \ell + 1} \big)
\nonumber
\\
{} & \quad\quad\quad\quad \times I_{\ell}^{+} (\bK_{1} , \bK_{2} , t) \bigg] ,
\label{CD_exp}
\end{align}
to stitch the piecewise
and analytical predictions.

Following all these manipulations,
for a given background bath, ${ n (\bK) }$,
a given pair of test particles, ${ (\bK_{1} , \bK_{2}) }$,
and a given initial angular separation, ${ \phi_{0} }$,
we have at our disposal a prediction for the time evolution
of the \PDF\ of the angular separation,
${ t \mapsto P (\phi \,|\, t) }$.
It is this \PDF\ that is used in the likelihood estimations.

\section{Numerical simulations}
\label{sec:NumericalSimulations}

In order to assess the validity of our approach,
it is important to test it on numerical simulations
of background baths of increasing complexity.

\subsection{Numerical simulations}
\label{sec:Nbody}

The numerical simulations were performed using
a multipole algorithm similar to the one presented
in~\cite{Fouvry+2022}, here adapted to the simpler case
of \VRR\@. Here, we only outline the main elements
of the algorithm.

All the radial orbit averages appearing in the coupling
coefficients $\mJ_{\ell}$ are replaced by discrete sampling
with $K$ nodes in eccentric anomaly.
The interaction Hamiltonian is truncated to the maximum
harmonic $\ellmax$.
Inspired by~\cite{Fouvry+2022},
the rates of change, ${ \rd \hbL / \rd t }$,
are computed in ${ \mO (N K \ellmax^{2}) }$ operations
owing to prefix sums.
We rewrite the equations of motion
as precession equations of the form ${ \rd \hbL / \rd t \!=\! \bO \!\times\! \hbL }$.
The dynamics
is integrated using the structure-preserving \texttt{MK2} scheme~\cite[see][]{Fouvry+2022}
with a constant timestep $h$.
This ensures the exact conservation of ${ |\hbL| \!=\! 1 }$
throughout the evolution.

\subsection{One-population bath}
\label{sec:SimpleFiducial}

Similarly to~\S{F} of~{G20},
we consider a single-population background bath.
Fixing the units to ${ G \!=\! \MBH \!=\! 1 }$,
we assume that it is composed of ${ N \!=\! 10^{3} }$ stars
of individual mass ${ m \!=\! 1 }$. Their distribution in $a$
is proportional to ${ a^{2 - \gamma} }$ with ${ \gamma \!=\! 1.5 }$,
and limited to ${ 1 \!\leq\! a \!\leq\! 100 }$. Finally,
their distribution in eccentricities is proportional to ${ e }$
and limited to ${ 0 \!\leq\! e \!\leq\! 0.3 }$.

For the test stars, we consider pairs of independent test particles
with ${ (a_{1} , e_{1}) \!=\! (9.5, 0.15) }$
and ${ (a_{2} , e_{2}) \!=\! (10.5, 0.05) }$.
For each realisation, we inject 500 pairs of test particles
with an initial angular separation set by ${ \phi_{0} \!=\! 3.0^{\circ}}$.
We performed a total of 500 realisations.
Following~\S\ref{sec:Nbody},
we used
${ K \!=\! 20 }$,
${ \ellmax \!=\! 10 }$,
${ h \!=\! 10^{-3}}$,
and integrated up to ${ \tmax \!=\! 100 }$.
Each realisation required ${ \sim 0.5 \, \rh }$ of computation
on two cores, with final relative errors in the total energy
(resp.\ total angular momentum)
of order $10^{-10}$ (resp.\ $10^{-10}$).

\subsection{Two-population bath}
\label{sec:LessSimpleFiducial}

We consider a more involved bath better mimicking SgrA*.
This is the Top-Heavy model from~\cite{Generozov+2020},
also reproduced in eq.~{(14)} of~\cite{Tep+2021}. We consider the same \BH\ mass
as in~\S\ref{sec:StellarCusps}. The bath is composed of two populations,
namely stars and \IMBHs\@, with
\begin{align}
\big[ \gamma_{\star} , m_{\star} , M_{\star} (\!<\! a_{0}) \big] {} & = \big[ 1.5 , 1 \, \Msun , 7.9 \!\times\! 10^{3} \, \Msun \big] ,
\nonumber
\\
\big[ \gamma_{\bullet} , m_{\bullet} , M_{\bullet} (\!<\! a_{0}) \big] {} & = \big[ 1.8 , 50 \, \Msun , 38 \!\times\! 10^{3} \, \Msun \big] ,
\label{TopHeavy}
\end{align}
fixing the scale radius to ${ a_{0} \!=\! 100 \, \mpc }$.
The distribution of eccentricities for both populations
is proportional to $e$
and limited to the range ${ 0 \!\leq\! e \!\leq\! 0.99 }$.

In practice, we can only simulate a finite number of background particles,
hence a finite range in semi-major axes.
The minimum (resp.\ maximum) semi-major axis
of the disc stars is ${ a_{\min}^{\star} \!\simeq\! 45 \, \mpc }$
(resp.\ ${ a_{\max}^{\star} \!\simeq\! 109 \, \mpc }$)
as given by S67 (resp.\ S87).
We fix the range of semi-major axes for the bath populations
to ${ a_{\min} \!\leq\! a \!\leq\! a_{\max} }$
with ${ a_{\min} \!=\! a_{\min}^{\star} / \eta }$,
${ a_{\max} \!=\! a_{\max}^{\star} \eta }$,
and ${ \eta \!=\! 5 }$.
The number of background stars
is then ${ N_{\star} \!=\! \!\int_{a_{\min}}^{a_{\max}} \!\! \rd a \, N (a) }$,
with ${ N(a) }$ following Eq.~\eqref{PDF_N}.
In practice, we find ${ N_{\star} \!=\! 83\,136 }$
and ${ N_{\bullet} \!=\! 5\,466 }$.

In each realisation, we inject a total of 100 independent discs
composed of 7 test stars with orbital parameters following~\S\ref{sec:StellarCusps}.
The disc stars' orientations are drawn from a Von Mises-Fischer \PDF\@ (see~\S{C2} in~{G20})
\begin{equation}
P (\hbL) \propto \re^{\kappa \hbL_{0} \cdot \hbL} 
\label{def_vMF}
\end{equation}
of concentration ${ \kappa \!=\! 2/(1 \!-\! \cos \phi_{0}) }$ with ${ \phi_{0} \!=\! 3^{\circ} }$,
and $\hbL_{0}$ the disc's mean orientation drawn uniformly on the unit sphere.
We performed a total of 500 realisations
using
${ K \!=\! 20 }$,
${ \ellmax \!=\! 10 }$,
${ h \!=\! 0.4 \, \kyr }$,
and integrated up to ${ \tmax \!=\! 10 \, \Myr }$.
Each realisation required ${ \sim 36 \, \rh }$ of computation
on two cores, with final relative errors in the total energy
(resp.\ total angular momentum)
of order $10^{-7}$ (resp.\ $10^{-8}$).

\section{Likelihood estimation}
\label{sec:Likelihood}

Ultimately, our goal is to constrain some parameters, $\bT$, for the background bath.
Let us first assume that we have at our disposal
one observed angular separation, $\phi_{ij}$, between two test stars.
We can write the likelihood of this angular separation given the bath parameters $\bT$ 
and the conserved orbital parameters $(\bK_{i},\bK_{j})$ as
\begin{equation}
P (\phi_{ij} \,|\, \bT, \bK_{i},\bK_{j})
= A \!\! \int \!\! \rd \mu \rd \sigma \, P (\phi_{ij} \,|\, \mu , \sigma) \, 
P (\mu , \sigma \,|\, \bT,\bK_{i},\bK_{j}) ,
\label{relation_Bayes_init}
\end{equation}
with $A$ a normalisation constant
and ${ P (\phi \,|\, \mu , \sigma) }$ the log-normal \PDF\
from Eq.~\eqref{lognormal_PDF}. In addition, we also introduced
\begin{equation}
P (\mu , \sigma \,|\, \bT) = \deltaD [\mu \!-\! \tmu (\bT , \bK_{i} , \bK_{j})] \, \deltaD [\sigma \!-\! \tsigma (\bT , \bK_{i} , \bK_{j})] ,
\label{PDF_model}
\end{equation}
with ${ \deltaD }$ the Dirac delta,
and ${ (\tmu (\bT , \bK_{i} , \bK_{j}) , \tsigma (\bT, \bK_{i} , \bK_{j}) ) }$ the kinetic predictions
of the log-normal parameters for some given bath,
as detailed in~\S\ref{sec:Lognormal}.
The likelihood $P (\phi_{ij} \,|\, \bT, \bK_{i},\bK_{j})$ is directly proportional
to the posterior probability $P (\bT \,|\, \phi_{ij}, \bK_{i},\bK_{j})$ derived from the 
measured parameters $(\phi_{ij}, \bK_{i},\bK_{j})$, provided we assume flat priors.
In this study, we use the likelihood as a proxy for the posterior measurements
of the bath's parameters.

\subsection{Marginalising over $\sigma$}
\label{sec:MarginalSigma}

In Fig.~\ref{fig:musigmaFiducial}, we noted that our prediction
for ${ \tsigma(\bT , \bK_{i} , \bK_{j}) }$ is poor. As a consequence, we marginalise over $\sigma$.
We write
\begin{equation}
P (\mu , \sigma \,|\, \bT , \bK_{i} , \bK_{j}) = P (\sigma \,|\, \mu) \, P (\mu \,|\, \bT, \bK_{i} , \bK_{j}) ,
\label{Bayes_marginal_sigma}
\end{equation}
with ${ P (\mu \,|\, \bT, \bK_{i} , \bK_{j}) \!=\! \deltaD [\mu \!-\! \tmu (\bT, \bK_{i} , \bK_{j})] }$
and assuming a flat prior for $\sigma$,
namely
\begin{equation}
P (\sigma \,|\, \mu) = \bone_{0 \leq \sigma \leq 1} ,
\label{prior_sigma}
\end{equation}
with $\bone$ the indicator function
inspired by the range in $\sigma$ effectively observed
in Fig.~\ref{fig:musigmaFiducial}.
Injecting Eq.~\eqref{Bayes_marginal_sigma} into Eq.~\eqref{relation_Bayes_init},
we can perform the integral over ${ \rd \mu \rd \sigma }$ to get
\begin{equation}
P (\phi_{ij} \,|\, \bT, \bK_{i} , \bK_{j}) = A \, I (\phi_{ij} \,|\, \tmu [\bT, \bK_{i} , \bK_{j}]) ,
\label{calc_int_sigma}
\end{equation}
where we introduced
\begin{equation}
I (\phi \,|\, \mu) \!=\! \!\! \int \!\! \rd \sigma \, P (\phi \,|\, \mu , \sigma) \, P (\sigma \,|\, \mu) \!=\! \frac{\Gamma [0 , \tfrac{1}{2} (\ln \phi \!-\! \mu)^{2}]}{2 \phi \, \sqrt{2 \pi}} ,
\label{calc_int_sigma_explicit}
\end{equation}
with ${ \Gamma [0 , x] }$ the incomplete gamma function.
If one has a sample ${ \{ \phi_{a} \}_{1 \leq a \leq n} }$ of distinct pairs of stars
at one's disposal, 
with the corresponding $(\bK_{i} , \bK_{j})$ implicit,
Eq.~\eqref{calc_int_sigma} becomes
\begin{equation}
P (\{ \phi_{a} \} \,|\, \bT ) = \prod_{a = 1}^{n} A \, I (\phi_a \,|\, \tmu [\bT]) ,
\label{calc_int_sigma_pop}
\end{equation}
i.e.\ pairs are treated as independent from one another,
and we marginalise over $\sigma$ for each.
The larger $P(\{ \phi_a \} \,|\, \bT)$,
the more likely the data for the given parameters.

We apply this method to the fiducial simulations
from~\S\ref{sec:SimpleFiducial},
considered at time ${ T_{\star} \!=\! 20 }$.
To obtain the likelihoods presented in Fig.~\ref{fig:CornerFiducial},
we fix the bath's parameters to their fiducial values (\S\ref{sec:SimpleFiducial})
except for ${ (m,\gamma) }$ or ${ (\phi_{0} , T_{\star}) }$,
which we respectively try to infer in the two panels of Fig.~\ref{fig:CornerFiducial}.
To mimick the observed data,
for some given parameters, we pick, at random, a total of 21 independent
pairs of simulated test stars whose angular separations, ${ \{ \phi_{a} \} }$,
are used in Eq.~\eqref{calc_int_sigma_pop}.
For every set of parameters, we repeat this procedure 100 times
and subsequently average the likelihood over this sample.
This has the effect of taking the expectation value of the likelihood
over an ensemble of realisations of the observations, and reduces the variation
due to the stochastic evolution of the pairs.
In that figure, the data was computed on a 500$\times$500 grid of parameters,
and subsequently smoothed with a Gaussian filter of standard deviation 6.
Finally, a contour labeled ${ x \% }$
corresponds to the level line ${ P (\bT) \!=\! p }$
where $p$ follows from ${ G (p) / G (p \!=\! 0) \!=\! x \% }$
with ${ G (p) \!=\! \!\int_{P (\bT) \geq p}\! \rd \bT P (\bT) }$.

\subsection{Dilution of a disc}
\label{sec:DiscDilution}

In practice, we are interested in the dilution of a disc composed of 7 stars
of individual orientations ${ \{ \hbL_{i} \}_{1 \leq i \leq 7} }$.
From these, we can construct ${ n \!=\! 21 }$ different pairs
of distinct stars of respective angular separations
\begin{equation}
\big\{ \phi_{a} \big\}_{1 \leq a \leq n} = \big\{ \cos^{-1} \big(\hbL_{i} \!\cdot\! \hbL_{j} \big) \big\}_{1 \leq i < j \leq 7} .
\label{pair_set}
\end{equation}
We use this sample of separations
in Eq.~\eqref{calc_int_sigma_pop}.

Let us now apply our likelihood estimation
to the simulations from~\S\ref{sec:LessSimpleFiducial},
which we consider at time ${ T_{\star} \!=\! 1.5 \, \Myr }$.
For the likelihoods shown in Fig.~\ref{fig:CornerSgrA},
we fix the bath's parameters to their fiducial values (\S\ref{sec:LessSimpleFiducial})
except for ${ (m_{\bullet} , \gamma_{\bullet}) }$
or ${ (\phi_{0} , T_{\star}) }$
(with ${ \kappa \!=\! 2/[1 \!-\! \cos \phi_{0}] }$),
which we respectively try to infer.
For a given set of parameters, we pick, at random, one simulated disc.
The initial conditions used in the prediction from~\S\ref{sec:NeighSep}
are obtained by drawing 7 stars according to Eq.~\eqref{def_vMF}
and following Eq.~\eqref{pair_set} to get the pairs' initial separation.
For every parameters, we repeat the procedure 100 times
and average ${ P (\{\phi_{a}\} \,|\, \bT) }$
for every pair over this sample,
again giving us the expected likelihood function
for this synthetic observation.
The associated inferences are illustrated in Fig.~\ref{fig:CornerSgrA},
using the same grid size and smoothing approach as in~\S\ref{sec:MarginalSigma}.

\subsection{Incorporating measurement errors}
\label{sec:MeasurementErrors}

When applying the present method to the observed S-stars,
we must also account for the measurement errors
in the stars' orbital parameters.
Fortunately, table~{3} of~\cite{Gillessen+2017}
provides us with measurement errors for the stars'
semi-major axes, eccentricities, inclinations and longitudes
of the ascending nodes.
For simplicity, we assume that
(i) these errors are independent
from one another;
(ii) to accommodate the physical ranges of each variables,
we assume that the semi-major axes follow Gamma distributions,
the eccentricities Beta distributions, and the inclinations and longitudes
Gaussian distributions,
with the reported means and variances.
For a given realisation,
the ``observed'' parameters ${ \{ \bK_{i} , \hbL_{i} \}_{1 \leq i \leq 7} }$
are drawn from these distributions,
and used in the likelihood estimation.
For each set of parameters, we repeat the procedure 100 times,
and average the likelihood over this sample.
This has the effect of marginalising the likelihood over the measurements
uncertainties, unlike the previous averaging procedures which produce the
expectation value of the likelihood for the simulated baths.
For SgrA* we have only a single realisation of the dynamics of the S-stars,
and this is a major source of the uncertainty in our inferences:
this can only be reduced by the inclusion of additional observed stars.

For the inferences presented in~\S\ref{sec:Results},
we systematically fix ${ (\gamma_{\star}, \gamma_{\bullet}) \!=\! (1.5, 1.8) }$
(see~\S\ref{sec:LessSimpleFiducial}),
as well as the age of the disc, ${ T_{\star} \!=\! 7.1 \, \Myr }$,
and the initial angular size of the disc set by ${ \phi_0 \!=\! 3^\circ }$.
In Fig.~\ref{fig:CornerSstar_massMenc},
we also fix ${ m_{\star} \!=\! 1\, \Msun }$ and the sum 
${ M_{\star}(\!<\! a_0) \!+\! M_{\bullet}(\!<\! a_0) \!=\! 45.9 \!\times\! 10^3 \, \Msun }$
for ${ a_{0} \!=\! 100 \, \mpc }$,
while varying $m_{\bullet}$ and the mass fraction of \IMBHs\@.
For Fig.~\ref{fig:CornerSstar_mratio}, 
we fix ${ [ M_{\star}(\!<\! a_0), M_{\bullet}(\!<\! a_0)] \!=\! [7.9 \!\times\! 10^3\, \Msun, 38 \!\times\! 10^3 \, \Msun] }$
and vary $m_{\star}$ and $m_{\bullet}/m_{\star}$.
Finally, we use the same grid size and smoothing procedure
as in~\S\ref{sec:MarginalSigma}.

\end{document}